\documentclass[runningheads]{llncs}

\usepackage{subfig}
\usepackage{adjustbox}
\usepackage{makeidx}
\usepackage{float}
\usepackage{multirow}
\usepackage{subfig}
\usepackage{graphicx}
\usepackage{booktabs}
\usepackage{amsfonts}       
\usepackage{amsmath}
\usepackage{amssymb}
\usepackage{amsthm}
\newtheorem{asu}{Assumption}
\usepackage{xcolor}
\usepackage[breaklinks=true,colorlinks,allcolors=blue]{hyperref}
\usepackage[symbol]{footmisc}

\usepackage{graphics}
\usepackage{nicefrac}       
\usepackage{microtype}      
\usepackage{wrapfig,lipsum,booktabs}
\usepackage{placeins}

\begin{document}
\title{Vector Quantisation for Robust Segmentation}
\titlerunning{Vector Quantisation for Robust Segmentation}
%
\footnotetext[1]{Joint first authors.}
\author{%
  Ainkaran Santhirasekaram\inst{1,*}
  \and
  Avinash Kori\inst{1,*}
  \and
  Mathias Winkler  \inst{2}
  \and 
  Andrea Rockall \inst{2}
  \and
  Ben Glocker \inst{1}
}
\institute{
  Department of Computing, Imperial College London, UK \and
  Department of Surgery and Cancer, Imperial College London, UK \\ \email{a.santhirasekaram19@imperial.ac.uk}
  }
\authorrunning{A Santhirasekaram et al.}
%
%

\maketitle   
\begin{abstract}

The reliability of segmentation models in the medical domain depends on the model's robustness to perturbations in the input space. Robustness is a particular challenge in medical imaging exhibiting various sources of image noise, corruptions, and domain shifts. Obtaining robustness is often attempted via simulating heterogeneous environments, either heuristically in the form of data augmentation or by learning to generate specific perturbations in an adversarial manner.  We propose and justify that learning a discrete representation in a low dimensional embedding space improves robustness of a segmentation model. This is achieved with a dictionary learning method called vector quantisation. We use a set of experiments designed to analyse robustness in both the latent and output space under domain shift and noise perturbations in the input space.  We adapt the popular UNet architecture, inserting a quantisation block in the bottleneck. We demonstrate improved segmentation accuracy and better robustness on three segmentation tasks. Code is available at \url{https://github.com/AinkaranSanthi/Vector-Quantisation-for-Robust-Segmentation}

\keywords{Robustness \and Vector Quantisation \and Semantic Segmentation.}
\end{abstract}
\section{Introduction}

Segmentation of medical images is important in both aiding diagnosis and treatment planning \cite{lei2020medical}. Deep learning, based on convolutional neural networks (CNNs), has significantly improved segmentation performance and is now the most widely used approach for automated segmentation \cite{lei2020medical}. However, it is well established in the literature that these models are not significantly robust to perturbations in the input whether that be noise or a domain shift \cite{carlini2017towards,moosavi2017universal}. This is particularly relevant in the medical domain whereby images are acquired from many sources with varying protocols and hence different image characteristics \cite{chen2021towards}. 
There have been various methods developed in the literature to increase robustness of the model most of which are based on simulating perturbations in the input space during training \cite{wiles2021fine}. For example, one can achieve this heuristically through various data augmentation strategies of the training data. One can also train a model to learn to generate data to have an adversarial effect on the performance of a model \cite{tramer2019adversarial,mummadi2019defending}.

The vector quantised variational auto-encoder (VQ-VAE) was proposed as a generative model which learns a discrete representation in the latent space via a method called vector quantisation \cite{van2017neural}. This is claimed to circumvent the issue of posterior collapse in the VAE \cite{van2017neural}. This model has been especially exploited in the field of image generation including text-to-image \cite{esser2021taming,gu2021vector,ding2021cogview,ramesh2021zero}.

We note that current methods have not explored how to improve the design of a segmentation model so that it is inherently more robust to input perturbations\cite{chen2021towards}. We take inspiration from the VQ-VAE and hypothesise that deep segmentation models are more robust and accurate when mapping the input data to a discrete latent space.

\subsection{Contribution}
We propose quantisation of the latent space of any segmentation network architecture, mapping the input images to a lower dimensional embedding space increasing robustness to perturbation in the input space. We provide a thorough justification for this claim under a set of laid out assumptions. We then derive an empirically driven upper bound for maximum allowed shift in the latent space due to perturbation for robustness to hold. We finally support our claim by demonstrating through a set experiments how robustness and performance of the popular UNet architecture \cite{ronneberger2015u} is improved with a quantised latent space. Our experiments look at two forms of perturbations to highlight our claim in the form of domain shift and noise.  We focus on anatomical segmentation which benefits most from a quantised latent space, because the spatial variability of human anatomy is structured and quantisation in the bottleneck aims to help to capture this by constraining the space where the features can reside.

\section{Methods}

\subsection{Robustness and Network Assumptions}
Given an input $x$, we first define a function $f(x)$ to represent the transformed input due to perturbation. This is a generic function in order to account for various types of perturbations ranging from a re-normalisation function to a non-linear mapping. We therefore now denote the perturbation to be $\delta(x) = f(x) - x $ which can represent noise or domain shift. The aim in this work is to find a way to learn a model ($\Phi$) with weights $w$ to be robust against $\delta(x)$ and construct an uncorrupted segmentation $y$ from the perturbed input $f(x)$. 

\begin{asu} 
\label{Robustness Assumption}
Assuming a small value for $\delta(x)$, we can then approximate $\Phi(x + \delta(x))$ with a first order Taylor expansion as follows: $\Phi(x + \delta(x)) = \Phi(x) + \delta(x)^T \nabla_x \Phi$.
Therefore, the training framework should optimize for $argmin_w [\Phi(x + \delta(x)) - \Phi(x)]$ to be robust.
\end{asu}

\begin{asu} 
\label{Network Assumption}
In this work we assume that the segmentation network can be decomposed into an encoder ($\Phi_e$) and decoder ($\Phi_d$) such that $\Phi = \Phi_d \circ \Phi_e$, where $\Phi_e: \mathcal{X} \rightarrow \mathcal{E}$ maps from image space to a lower dimensional embedding space and $\Phi_d: \mathcal{E} \rightarrow \mathcal{Y}$ maps the embedding space back to segmentation space. 
 
\end{asu}

\subsection{Quantisation for Robustness}

Formally, with the quantisation block our segmentation network $\Phi$ now decomposes as $\Phi_d \circ \Phi_q \circ \Phi_e$, where $\Phi_e, \Phi_d, \Phi_q$ corresponds to the encoder, decoder, and quantisation blocks.  $\Phi_q$ maps the embedding vectors ($e$) from the continuous embedding space output of $\Phi_e(x)$ to quantised vectors ($z_q$). 
The goal of the quantisation block is to remove unnecessary information in the latent space by collapsing a continuous latent space to a set of discrete vectors.

The quantisation process initially requires us to define a codebook ($c \in \mathcal{R}^{K \times D}$. $K$ is the size of the codebook and $D$ is the dimensionality of each codebook vector $l_i \in \mathcal{R^D}$. We then define a discrete uniform prior and learn a categorical distribution $\mathbb{P}(z \mid x)$ with one-hot probabilities determined by the mapping of each embedding vector in $e$ to the nearest codebook vector $l_k$ which form $z_q$ as follows \cite{van2017neural}:

\begin{equation}
  \label{eq:1}
  \mathbb{P}(z = k\mid x) = 
    \begin{cases}
      1,      &   \text{for} \quad k = argmin_i || \Phi_e(x) - l_i ||_2 \\
      0,      &   \text{otherwise}
    \end{cases}
\end{equation}

Backpropagation through the non-differentiable quantisation block requires straight-through gradient approximation whereby one copies the gradients from $z_q$ to the encoder output ($e$) which is used to update the codebook. This allows the entire model to be trained end-to-end with the following loss function\cite{van2017neural}:
\begin{equation}
  \label{eq:2}
  \mathcal{L}_{total} = \mathcal{L}_{Dice}(\hat{y},y) + \mathcal{L}_{CE}(\hat{y},y) + \|sg(\Phi_e(x))-l\|_2 + \beta\|\Phi_e(x)-sg(l)\|_2 \
\end{equation}

The first two terms in equation \ref{eq:2} correspond to the Dice and cross entropy loss between the predicted segmentation ($\hat{y}$) and label ($y$). The third term updates the codebook by moving the codebook vectors $(l_i)$ towards the output of the encoder.
The fourth term in equation \ref{eq:2} is defined as a commitment loss weighted by $\beta$ \cite{van2017neural} . 
A stop gradient (sg) is applied to constrain the update to the appropriate operand.

Based on assumption \ref{Robustness Assumption}, we get, $\Phi_q(\Phi_e(x + \delta(x))) = \Phi_q(\Phi_e(x) + \delta(x)^T \nabla_x \Phi_e(x)) $. 

We claim, quantisation pushes $\delta(x)^T \nabla_w \Phi_e(x)$ to 0 and thereby enforces $\Phi_q(\Phi_e(x + \delta(x))) = \Phi_q(\Phi_e(x))$. This claim holds true, if we make the following assumption:

\begin{asu} 
\label{Quantization Assumption}

We assume if $\|\Phi_e(x)- l_i\|_2 > 0$; then x is absolutely perturbed by $\delta(x)$. This means a codebook $c$ with dimensionality $D$ contains the minimal number of codebook vectors $K$ to fully capture all possible semantics in the latent space i.e., complete. We also assume $c$ is uniformly distributed on the surface of a D-dimensional hypersphere. Therefore, the space on the hypersphere which lie between $c$ represents only perturbations of $c$. We denote the entire surface of the hypersphere as $Z$ and $\Phi_e(x)$ only generates $e$ which only lies on $Z$. 
\end{asu}

Finally, if the decoder ($\Phi_d$) is only a function of the quantised representation ($z_q$) then given our assumption \ref{Quantization Assumption}, $\Phi(x + \delta(x)) = \Phi(x)$. However, if $\Phi_d$ is a function of $z$ and output of each scale from the encoder ($s$) like in the UNet, then the effect on the output of the model by $\delta(x)$ is only reduced. Yet, this maybe beneficial in practise where the codebook is not complete.

\subsection{Perturbation Bounds}
A codebook has the advantage to allow us to derive a limit for the shift in latent space which represents the boundary between perturbation and a true semantic shift for the data distribution which we sample, given assumption \ref{Quantization Assumption}. This can be defined as the maximum perturbation allowed around a single codebook vector denoted $r$ and calculated empirically as half the average distance between a codebook vector ($l_i$) and its nearest neighbour ($l_{i + 1}$) across the whole of $c$ as follows:

\begin{equation}
  \label{eq:3}
  r = \dfrac{\sum_{i=0}^{i=K - 1} \frac{1}{2}(\| l_i -  l_{i+1}\|_{2})}{K-1}\\
\end{equation}

Uniformity also allows to state no matter what the shift along the surface of $Z$, one will always be at least a distance $r$ from the closest codebook vector $l_k$.

Next, for simplicity observe a single vector from the output of $\Phi_e(x)$ and $\Phi_e(x + \delta(x))$  denoted $e_j$ and $e_j + \Delta$. We can combine equation \ref{eq:3} and the first order Taylor expansion of $\Phi_e(x + \delta(x))$ to theoretically express r in terms of $\delta(x)$ as follows: 
\begin{equation}
  \label{eq:4}
    r >  \|\delta(x)^T\nabla_x e_j\|_2
\end{equation}

Therefore to affect an output of the quantisation block $\Phi_q$, a perturbation $\delta(x)$ should lead to a change in the embedding space ($e$) greater than $r$ whose upper bound expressed in terms of $\delta(x)$ is derived in equation \ref{eq:4}. 

\subsection{Implementation Details and Data}

\subsubsection{Architecture:}
We consider the UNet as our benchmark segmentation architecture and for the proposed architecture, \textit{VQ-UNet}, we add a vector quantisation block at the bottleneck layer of the baseline UNet. Our codebook size ($K$) is 1024 each of dimension ($D$) 256. We consider both 2D and 3D UNets. In the encoder we double the number of feature channels from 32 and 16 at the first level to 512 and 256 at the bottleneck, respectively for 2D and 3D. Each scale of the encoder and decoder consist of a single pre-activation residual block \cite{he2016identity}, with group normalisation (\cite{wu2018group}) and Swish activation (\cite{ramachandran2017searching}).

\subsubsection{Training:}

We fine-tuned the hyper-parameter $\beta$ in the loss function equation \ref{eq:3} to be 0.25. The loss function for training the UNet is the sum of the first two terms of equation \ref{eq:2} (Dice/cross entropy). We train with batch-size of 10 and 2 for the 2D and 3D tasks, respectively. We apply the same spatial augmentation strategy for all models, and use Adam optimisation with a base learning rate of 0.0001  and  weight  decay of 0.05\cite{kingma2014adam}. We train all models for a maximum of 500 epochs on three NVIDIA RTX 2080 GPUs.

\subsubsection{Datasets:} We use the following three datasets for our experimental study:

\textit{Abdomen:} We use the Beyond the Cranial Vault (BTCV) consisting of 30 CT scans with 13 labels acquired from a single domain (Vanderbilt University Medical Center) \cite{landman2015miccai}. All images were normalised to 0-1 and resampled to 1.5$\times$1.5$\times$2mm. We randomly crop 96$\times$96$\times$96 patches for training. 

\textit{Prostate:} The prostate dataset originates from the NCI-ISBI13 Challenge \cite{bloch2015nci}. It consists of 60 T2 weighted MRI scans of which half come from Boston Medical Centre (BMC) acquired on a 1.5T scanner with an endorectal coil and the other half is acquired from Radboud University Nijmegen Medical Centre (RUNMC) on a 3T scanner with a surface coil \cite{bloch2015nci}. All images were re-sampled to 0.5$\times$0.5$\times$1.5mm and z-score normalized. We centre crop to 192$\times$192$\times$64.

\textit{Chest-X-ray:} We use the NIH Chest X-ray dataset \cite{wang2017chestx} with annotations provided by \cite{tang2019xlsor} and the Japanese Society of Radiological Technology (JSRT) dataset \cite{shiraishi2000development} for domain shift analysis; there are 100 and 154 annotated images, respectively. Images were resized to 512$\times$512 pixels and normalised to 0-1.

\section{Experiments}
\subsection{Codebook Study}
We first analyse whether assumption \ref{Quantization Assumption} holds by calculating $r$ based on equation \ref{eq:3} and its standard deviation. We note there is a very large standard deviation around $r$ ranging from 0.0011 to 0.0021 for all 5 datasets (Table~\ref{tab2}). This suggests the 5 codebooks are not uniformly distributed i.e.,  incomplete. Hence, we cannot reliably assume that a shift in latent space greater than $r$ represents the boundary between a meaningful semantic shift and perturbation. Therefore, $r$ is obsolete, and we can only denote $r_i$; the distance for each codebook vector ($l_i$) to its nearest neighbour.  $r_i$ allows us to at least represent the maximally allowed perturbation in the latent space for each learnt codebook vector.

\begin{table}
\centering
\caption{Mean $r \pm 1$ standard deviation  for all 5 datasets}.
\label{tab1}
\begin{tabular}{ccccc}
\hline
\multicolumn{1}{|c|}{\textbf{NIH}}&
\multicolumn{1}{|c|}{\textbf{JRST}}         
& \multicolumn{1}{c|}{\textbf{Abdomen}} & \multicolumn{1}{c|}{\textbf{BMC}} & \multicolumn{1}{c|}{\textbf{RUNMC}}
\\ \hline
\multicolumn{1}{|c|}{0.001 $\pm$ 0.011} & \multicolumn{1}{c|}{0.002 $\pm$ 0.014} &
\multicolumn{1}{c|}{0.011 $\pm$ 0.018} & \multicolumn{1}{c|}{0.015 $\pm$ 0.021} &      \multicolumn{1}{c|}{0.012 $\pm$ 0.019} 
\\ \hline
\end{tabular}
\end{table}

\subsection{Domain Shift Study} 
We tackle domain shift from the angle of model design through incorporation of a vector quantisation block in the  UNet bottleneck. We evaluate how segmentation performance of the VQ-UNet differs from the UNet on a single domain and across domain for the chest X-ray and prostate datasets on two evaluation metrics: : Dice score and $95\%$ Hausdorff distance in mm (HD95). We randomly split a single domain in the prostate dataset into 24 for training and 6 for validation and use the best trained model based on the Dice score for testing on the second domain (30). For the NIH and JRST chest X-ray datasets, we randomly select 20 and 30 samples respectively for validation to find the best model to test on the second domain.

\begin{table}[t]
\centering
\caption{Mean Dice and HD95 on the validation sets for a single domain and test set across domain. The arrow represents the domain shift }.
\label{tab2}
\begin{tabular}{cccccccccc}
\hline
\multicolumn{9}{|c|}{\textbf{Chest X-ray}}
\\ \hline
\multicolumn{1}{|c|}{}               & \multicolumn{2}{c|}{\textbf{JRST}} & \multicolumn{2}{c|}{\textbf{NIH}} & \multicolumn{2}{c|}{\textbf{JRST$\rightarrow$NIH}} & \multicolumn{2}{c|}{\textbf{NIH$\rightarrow$JRST}}  \\ \cline{2-9}
\multicolumn{1}{|c|}{}               &
\multicolumn{1}{c|}{\textbf{Dice}} & \multicolumn{1}{c|}{\textbf{HD95}} &
\multicolumn{1}{c|}{\textbf{Dice}} & \multicolumn{1}{c|}{\textbf{HD95}} &      \multicolumn{1}{c|}{\textbf{Dice}} & \multicolumn{1}{c|}{\textbf{HD95}} &
\multicolumn{1}{c|}{\textbf{Dice}} & \multicolumn{1}{c|}{\textbf{HD95}} 
\\ \hline
\multicolumn{1}{|c|}{\textbf{UNet}} &
\multicolumn{1}{c|}{0.93} & \multicolumn{1}{c|}{7.31} &
\multicolumn{1}{c|}{0.96} & \multicolumn{1}{c|}{6.80} &      \multicolumn{1}{c|}{0.95} & \multicolumn{1}{c|}{7.12} &
\multicolumn{1}{c|}{0.82} & \multicolumn{1}{c|}{8.27} 
\\ \hline
\multicolumn{1}{|c|}{\textbf{VQ-UNet}} &
\multicolumn{1}{c|}{\textbf{0.94}} & \multicolumn{1}{c|}{\textbf{7.21}} &
\multicolumn{1}{c|}{\textbf{0.970}} & \multicolumn{1}{c|}{\textbf{6.01}} &      
\multicolumn{1}{c|}{\textbf{0.96}} & \multicolumn{1}{c|}{\textbf{6.51}} &
\multicolumn{1}{c|}{\textbf{0.85}} & \multicolumn{1}{c|}{\textbf{7.79}} 
\\ \hline

\multicolumn{9}{|c|}{\textbf{Prostate}}
\\ \hline
\multicolumn{1}{|c|}{}               & \multicolumn{2}{c|}{\textbf{BMC}} & \multicolumn{2}{c|}{\textbf{RUNMC}} & \multicolumn{2}{c|}{\textbf{BMC$\rightarrow$RUNMC}} & \multicolumn{2}{c|}{\textbf{RUNMC$\rightarrow$BMC}}  \\ \cline{2-9}
\multicolumn{1}{|c|}{}               &
\multicolumn{1}{c|}{\textbf{Dice}} & \multicolumn{1}{c|}{\textbf{HD95}} &
\multicolumn{1}{c|}{\textbf{Dice}} & \multicolumn{1}{c|}{\textbf{HD95}} &      \multicolumn{1}{c|}{\textbf{Dice}} & \multicolumn{1}{c|}{\textbf{HD95}} &
\multicolumn{1}{c|}{\textbf{Dice}} & \multicolumn{1}{c|}{\textbf{HD95}} 
\\ \hline
\multicolumn{1}{|c|}{\textbf{UNet}} &
\multicolumn{1}{c|}{0.80} & \multicolumn{1}{c|}{8.42} &
\multicolumn{1}{c|}{\textbf{0.824}} & \multicolumn{1}{c|}{7.84} &
\multicolumn{1}{c|}{0.55} & \multicolumn{1}{c|}{33.3} &
\multicolumn{1}{c|}{0.62} & \multicolumn{1}{c|}{25.7} 
\\ \hline
\multicolumn{1}{|c|}{\textbf{VQ-UNet}} &
\multicolumn{1}{c|}{\textbf{0.82}} & \multicolumn{1}{c|}{\textbf{7.82}} &
\multicolumn{1}{c|}{0.822} & \multicolumn{1}{c|}{\textbf{7.11}} &      \multicolumn{1}{c|}{\textbf{0.59}} & \multicolumn{1}{c|}{\textbf{31.5}} &
\multicolumn{1}{c|}{\textbf{0.71}} & \multicolumn{1}{c|}{\textbf{21.4}}
\\ \hline
\end{tabular}
\end{table}

\begin{figure}[h]
    \centering
\includegraphics[trim={0.5cm 6.5cm 0 4.0cm},clip, scale=0.41]{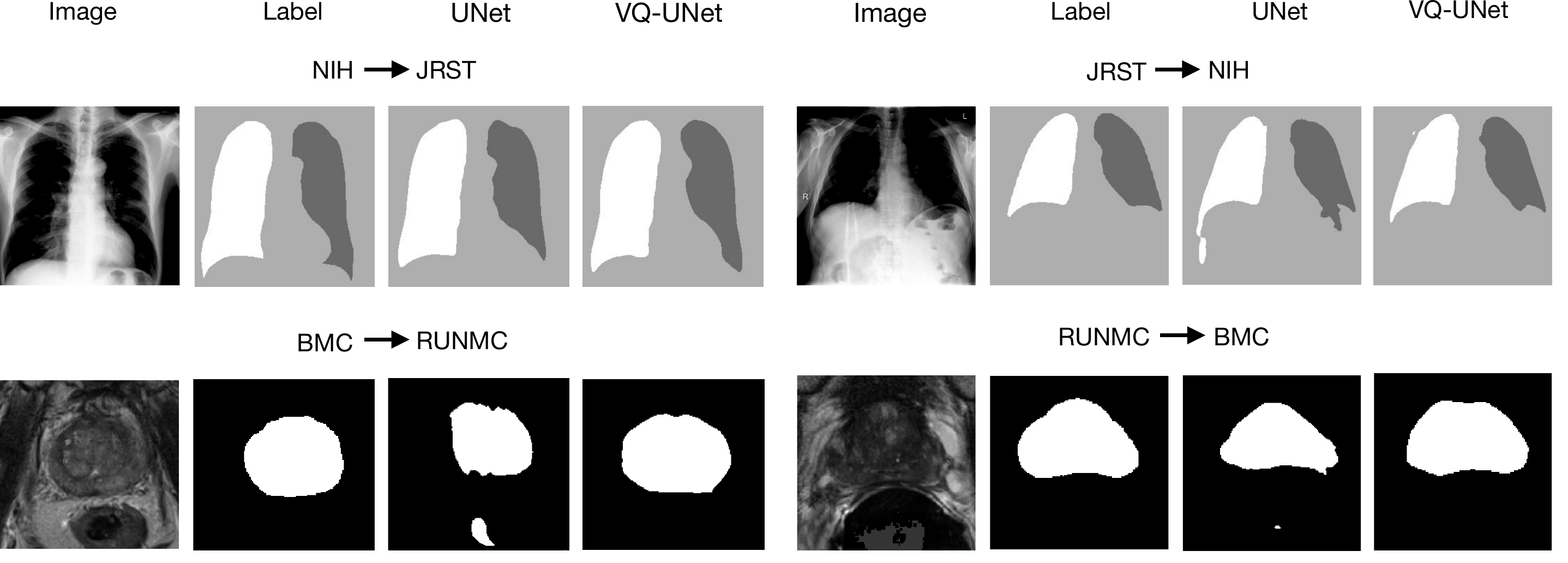}
\caption{Sampled image input and Segmentation output for 2 domain shifts in chest X-ray (top row) and prostate (bottom row)}.
\label{fig2}
\end{figure}

Overall, the VQ-UNet improved the segmentation performance both on the validation set and test set from a different domain for both prostate and chest X-ray (Table~\ref{tab2}). We note the UNet Dice score reduces to 0.82 and 0.93 compared to 0.85 from 0.94 for the VQ-UNet when shifting domain from JRST to NIH (Table~\ref{tab2}). For prostate, there is a significant domain shift and we note a significant drop in Dice score and HD95 distances when testing on a different domain for both the UNet and VQ-UNet (Table~\ref{tab2}). However, VQ-UNet appears to be more robust to this domain shift. This is particularly noted when testing on the BMC dataset after training the VQ-UNet on RUNMC (Table~\ref{tab2}).  The drop in performance albeit improved compared the UNet, is due to an incomplete codebook. It is highly likely the data from the test set maps to $e$ which is a distance greater than $r_i$ of the nearest codebook vector ($l_k$). This suggest $e$ is a perturbed version of a discrete point on the hypersphere which is not in our incomplete codebook. 
Nonetheless, in Fig~\ref{fig2} we note the smoother, anatomically more plausible segmentation map of the VQ-UNet compared to the UNet. 

\subsection{Perturbation Study}
We compare how much the latent space changes in both models with different perturbations in the input space for three datasets (abdomen, NIH, BMC). There are myriad of perturbations one can apply in the input space, so we choose three different types of noise perturbations (Gaussian, salt and pepper, and Poisson noise) under 5 noise levels ranging from $0\%$ to $30\%$ to justify our claim of robustness.  

\begin{figure*}
    \centering
    \includegraphics[trim={0.5cm 5.5cm 0 5.5cm},clip, scale=0.41]{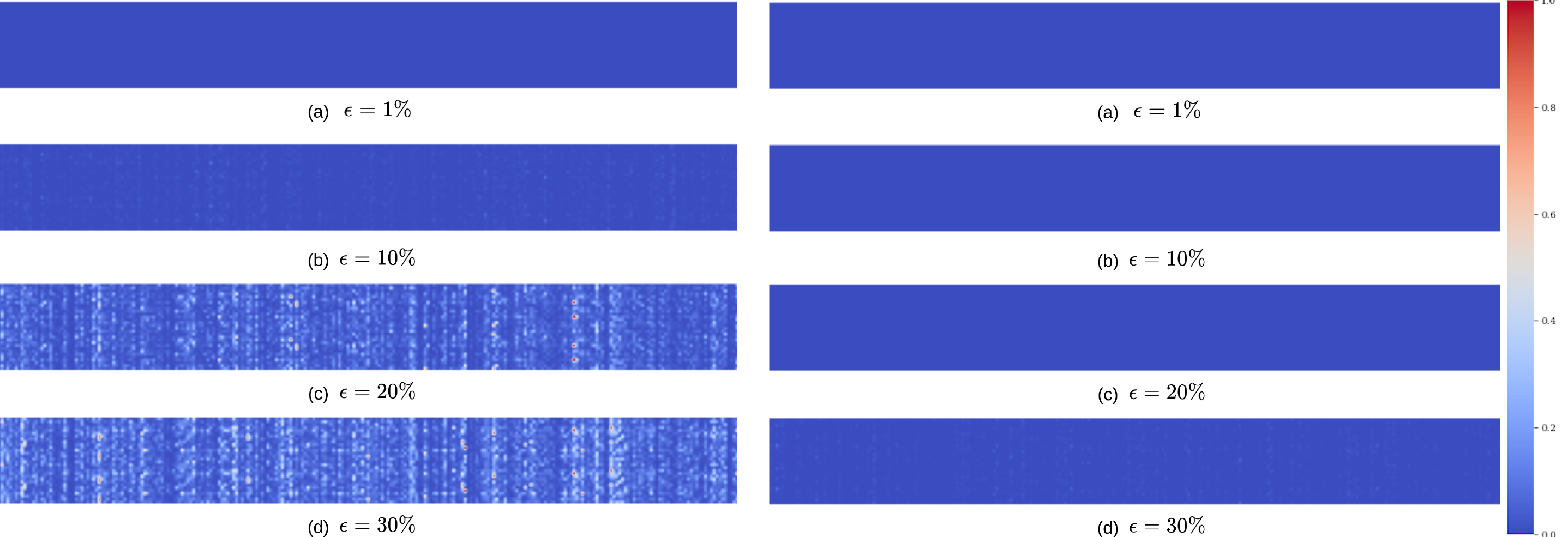}
    \caption{Variance heatmap of UNet(left) and VQ-UNet(right) latent space under 4 Gaussian noise levels for the NIH dataset. X-axis indicates a unique subset of features from a latent space, Y-axis corresponds to 100 randomly sampled test set images, and value at each location indicates the variance of a specific feature for a given image across 100 test time augmentations with Gaussian noise.}
    \label{fig:varianceheatmap}
\end{figure*}

To evaluate the effect of noise on the latent space of the trained models, we sample $100$ different noise vectors for each image at each noise level, and observe the variance in the latent space on the validation set.
Table \ref{tab:LatentVariation} describes the average variance of latent space features in both models across all noise levels for each type of noise. It can seen that latent space features in VQ-UNet are not significantly changed (close to 0 variance) under various types of noise. The results are visualised in Fig. \ref{fig:varianceheatmap} whereby the latent space of the VQ-UNet does not significantly change compared to the UNet under the addition of up to $30\%$ Gaussian noise in the NIH dataset. Therefore, given equation \ref{eq:4}, noise levels of up to $30\%$ is leading to a shift in the latent space of the VQ-UNet less than $r_i$.

\begin{table*}
\centering
\caption{Average latent space variance in both the models for all three datasets.}
\label{tab:LatentVariation}
\begin{tabular}{|c|ccc|ccc|ccc|}
\hline
              & \multicolumn{3}{c|}{\textbf{Abdominal CT}}  
              & \multicolumn{3}{c|}{\textbf{Chest X-ray}}   
              & \multicolumn{3}{c|}{\textbf{Prostate}}      \\ \hline
              & \multicolumn{1}{c|}{\textbf{\begin{tabular}[c]{@{}c@{}}Gauss.\\ Noise\end{tabular}}} & \multicolumn{1}{c|}{\textbf{\begin{tabular}[c]{@{}c@{}}S \&P\\ Noise\end{tabular}}} & \textbf{\begin{tabular}[c]{@{}c@{}}Poisson\\ Noise\end{tabular}} & \multicolumn{1}{c|}{\textbf{\begin{tabular}[c]{@{}c@{}}Gauss.\\ Noise\end{tabular}}} & \multicolumn{1}{c|}{\textbf{\begin{tabular}[c]{@{}c@{}}S \&P\\ Noise\end{tabular}}} & \textbf{\begin{tabular}[c]{@{}c@{}}Poisson\\ Noise\end{tabular}} & \multicolumn{1}{c|}{\textbf{\begin{tabular}[c]{@{}c@{}}Gauss.\\ Noise\end{tabular}}} & \multicolumn{1}{c|}{\textbf{\begin{tabular}[c]{@{}c@{}}S \&P\\ Noise\end{tabular}}} & \textbf{\begin{tabular}[c]{@{}c@{}}Poisson\\ Noise\end{tabular}} \\ \hline
\textbf{UNet} & \multicolumn{1}{c|}{0.46}                                                            & \multicolumn{1}{c|}{0.44}                                                           & 0.46                                                             & \multicolumn{1}{c|}{0.51}                                                            & \multicolumn{1}{c|}{0.43}                                                           & 0.47                                                             & \multicolumn{1}{c|}{0.56}                                                            & \multicolumn{1}{c|}{0.51}                                                           & 0.51                                                             \\ \hline
\textbf{VQ-UNet} & \multicolumn{1}{c|}{\textbf{3e-4}}                                                            & \multicolumn{1}{c|}{\textbf{5e-5}}                                                           & \textbf{2e-4}                                                             & \multicolumn{1}{c|}{\textbf{2e-4}}                                                            & \multicolumn{1}{c|}{\textbf{1e-4}}                                                           & \textbf{3e-4}                                                             & \multicolumn{1}{c|}{\textbf{1e-4}}                                                            & \multicolumn{1}{c|}{\textbf{6e-5}}                                                           & \textbf{8e-5}                                                             \\ \hline
\end{tabular}
\end{table*}

\begin{figure}[h]
    \centering
\includegraphics[trim={4.5cm 11.0cm 2.5cm 6.0cm},clip, scale=0.52]{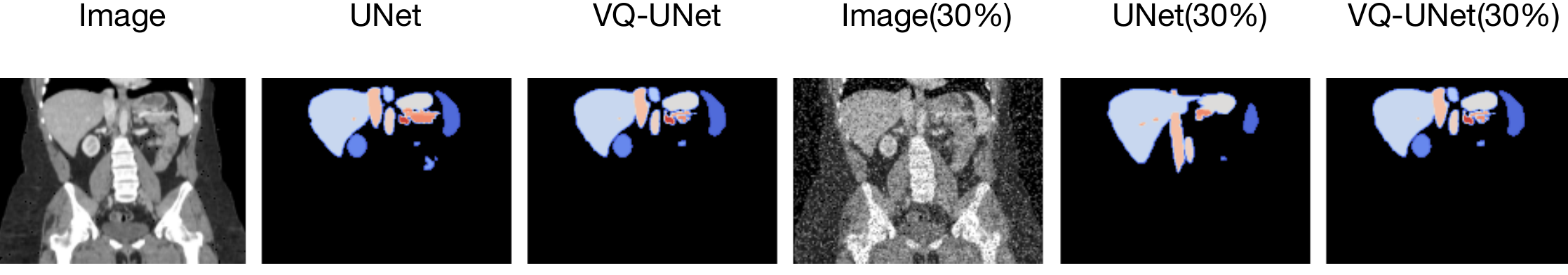}
\caption{Sampled Abdomen input image and Segmentation output for UNet and VQ-UNet under $0\%$ (1$^{st}$ 3 columns) and $30\%$ (2$^{nd}$ 3 columns) for s\&p noise}.
\label{fig:noiserecon}
\end{figure}

In our analysis of the output space, table \ref{tab:GaussianResults} indicates the effect of Gaussian perturbation on Dice scores on the in-domain validation set. It demonstrates the Dice scores are more stable in the VQ-UNet compared to the UNet for all three datasets up to $30\%$ noise. We highlight this result further in Figure \ref{fig:noiserecon} which demonstrates that the segmentation maps produced by the VQ-UNet under the addition of $30\%$ Gaussian noise do not change visually compared to the UNet. We make similar findings for salt \& pepper noise and Poisson noise (see supplementary material).

\begin{table*}[b]
\centering
\caption{Gaussian noise perturbation on all 3 datasets}.
\label{tab:GaussianResults}
\begin{tabular}{|cccccc|}
\hline

\multicolumn{1}{|c|}{}               & \multicolumn{1}{c|}{\textbf{Dice @0\% }} & \multicolumn{1}{c|}{\textbf{Dice  @1\%  }} & \multicolumn{1}{c|}{\textbf{Dice  @10\%  }} & \multicolumn{1}{c|}{\textbf{Dice  @20\% }} & \textbf{Dice  @30\%  } \\ \hline
\multicolumn{6}{|c|}{\textbf{Chest X-ray NIH dataset}}  \\ \hline
\multicolumn{1}{|c|}{\textbf{UNet}}  & 
\multicolumn{1}{c|}{0.95 $\pm$ 0.02}                               & 
\multicolumn{1}{c|}{0.96 $\pm$ 0.02}                               & \multicolumn{1}{c|}{0.95 $\pm$ 0.03}                                & \multicolumn{1}{c|}{0.95 $\pm$ 0.03}                                &            
0.95 $\pm$ 0.03                     \\ \hline
\multicolumn{1}{|c|}{\textbf{VQ-UNet}} & 
\multicolumn{1}{c|}{\textbf{0.97 $\pm$ 0.01}}                               & \multicolumn{1}{c|}{\textbf{0.97 $\pm$ 0.01}}                               & \multicolumn{1}{c|}{\textbf{0.97 $\pm$ 0.01}}                                & \multicolumn{1}{c|}{\textbf{0.96 $\pm$ 0.02}}                                &        
\textbf{0.96 $\pm$ 0.02}                         \\ \hline
\multicolumn{6}{|c|}{\textbf{Abdominal CT }}  \\ \hline

\multicolumn{1}{|c|}{\textbf{UNet}}  & 
\multicolumn{1}{c|}{0.77 $\pm$ 0.01}                               & 
\multicolumn{1}{c|}{0.76 $\pm$ 0.02}                               & 
\multicolumn{1}{c|}{0.77 $\pm$ 0.04}                                & 
\multicolumn{1}{c|}{0.76 $\pm$ 0.04}                                &    
0.75  $\pm$ 0.08 \\ \hline
\multicolumn{1}{|c|}{\textbf{VQ-UNet}} & 
\multicolumn{1}{c|}{\textbf{0.80 $\pm$ 0.01}}                               & 
\multicolumn{1}{c|}{\textbf{0.79 $\pm$ 0.01}}                               & 
\multicolumn{1}{c|}{\textbf{0.80 $\pm$ 0.01}}                               & 
\multicolumn{1}{c|}{\textbf{0.80 $\pm$ 0.02}}                               &      
\textbf{0.79 $\pm$ 0.02}                       \\ \hline
\multicolumn{6}{|c|}{\textbf{Prostate BMC dataset}} \\ \hline

\multicolumn{1}{|c|}{\textbf{UNet}}  & 
\multicolumn{1}{c|}{0.80  $\pm$ 0.02}                               & 
\multicolumn{1}{c|}{0.81  $\pm$ 0.02}                               & 
\multicolumn{1}{c|}{0.80  $\pm$ 0.03}                                & 
\multicolumn{1}{c|}{0.78 $\pm$ 0.03}                                &        
0.77     $\pm$ 0.06                     \\ \hline
\multicolumn{1}{|c|}{\textbf{VQ-UNet}} & 
\multicolumn{1}{c|}{\textbf{0.82 $\pm$ 0.02}}                               & 
\multicolumn{1}{c|}{\textbf{0.82 $\pm$ 0.02}}                               & 
\multicolumn{1}{c|}{\textbf{0.82 $\pm$ 0.02}}                                & 
\multicolumn{1}{c|}{\textbf{0.82 $\pm$ 0.03}}                                &            
\textbf{0.80     $\pm$ 0.04}                 \\ \hline
\end{tabular}
\end{table*}

Overall, the perturbation experiments show that quantisation helps in mitigating the effect of noise perturbation on the latent space, thereby generating non-corrupted segmentation maps. This is in contrast to the prostate domain shift experiments whereby the domain shift generates a shift in the latent space larger than $r_i$ for each codebook vector or maps to perturbations from discrete points not present in the codebook. 

\section{Conclusion}

We propose and justify that given a segmentation architecture which maps the input space to a low dimensional embedding space, learning a discrete latent space via quantisation improves robustness of the segmentation model. We highlight quantisation to be especially useful in the task of anatomical segmentation where the output space is structured and hence the codebook metaphorically behaves like an atlas in latent space. This however also possibly limits quantisation in highly variable segmentation tasks such as tumour segmentation. 

For future work, other architectures under various other perturbations such as adversarial perturbations will be explored. We also note the limitation of having a uniform prior during training in this work and aim to further increase robustness by jointly training a VQ model with an auto-regressive prior. 

\noindent \textbf{Acknowledgements.} This work was supported and funded by Cancer Research UK (CRUK) (C309/A28804) and UKRI centre for Doctoral Training in Safe and Trusted AI (EP/S023356/1).

\bibliographystyle{splncs04}
\bibliography{main.bib}

\section*{\centering Supplementary Material}              
\subsection*{Derivation of equation 4:} 
\begin{align*}
    \begin{split}
    \quad  r &= \frac{\sum_{i=0}^{i=K - 1} \frac{1}{2}(\| l_i -  l_{i+1}\|_{2})}{K-1}  \quad \text{:Equation 3}\\
    e_i - (e_i + \Delta) &=  \delta(x)^T\nabla_x e_i \quad \text{:First order Taylor expansion of $\Phi_e(x + \delta(x))$} \\
    \|e_i - (e_i + \Delta)\|_2 &=  \|\delta(x)^T\nabla_x e_i\|_2 \\
    & \therefore r >  \|\delta(x)^T\nabla_x e_j\|_2 \\
    & x, \delta(x) \in \mathcal{R}^{N} \quad e_i, e_i + \delta \in \mathcal{R}^{D}   \quad  \nabla_x e_i \in \mathcal{R}^{N\times D} 
    \end{split}
\end{align*}

\subsection*{Figures and Tables}
 \begin{figure}
     \centering
 \includegraphics[trim={0.5cm 2.9cm 0 3.95cm},clip, scale=0.4]{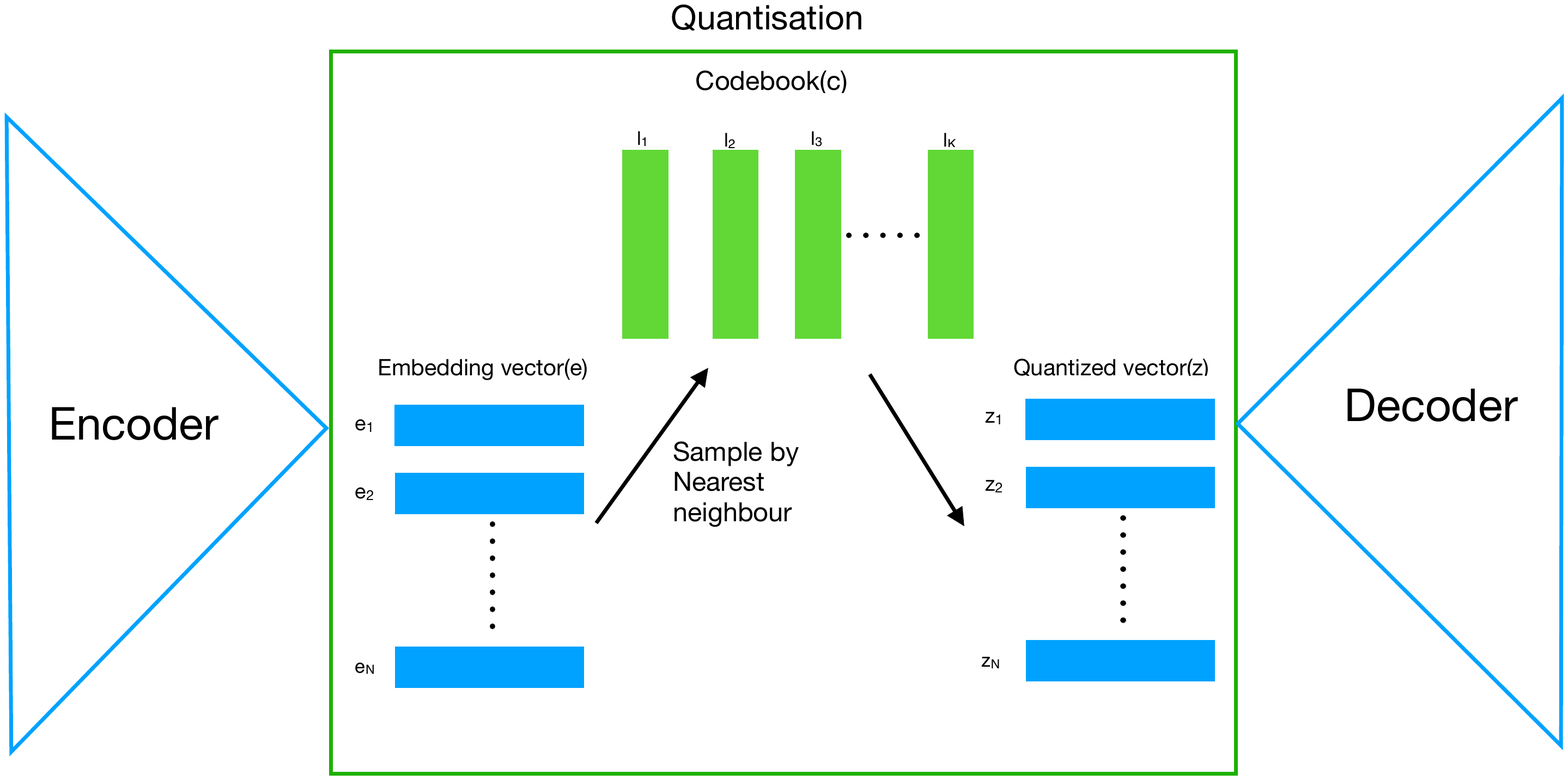}
 \caption{Proposed vector quantisation of the bottleneck of segmentation model. Note, skip connections between the encoder and decoder are optional}.
 \label{fig1}
 \end{figure}
\begin{table}[]
\centering
\caption{Salt and Pepper perturbation results on all 3 datasets.}
\label{tab:snpResults}
\begin{tabular}{|cccccc|}
\hline
\multicolumn{1}{|c|}{}               & \multicolumn{1}{c|}{\textbf{Dice @0\% }} & \multicolumn{1}{c|}{\textbf{Dice @1\% }} & \multicolumn{1}{c|}{\textbf{Dice @10\%}} & \multicolumn{1}{c|}{\textbf{Dice @20\% }} & \textbf{Dice @30\% } \\ \hline
\multicolumn{6}{|c|}{\textbf{Chest X-ray NIH dataset}} \\ \hline
\multicolumn{1}{|c|}{\textbf{UNet}}  & 
\multicolumn{1}{c|}{0.96 $\pm$ 0.02}                               & 
\multicolumn{1}{c|}{0.95 $\pm$ 0.02}                               & 
\multicolumn{1}{c|}{0.92 $\pm$ 0.05}                                & 
\multicolumn{1}{c|}{0.89 $\pm$ 0.08}                                &            
0.86 $\pm$ 0.08                     \\ \hline
\multicolumn{1}{|c|}{\textbf{VQ-UNet}} & 
\multicolumn{1}{c|}{\textbf{0.97 $\pm$ 0.01}}                               & \multicolumn{1}{c|}{\textbf{0.97 $\pm$ 0.01}}                               & \multicolumn{1}{c|}{\textbf{0.95 $\pm$ 0.01}}                                & \multicolumn{1}{c|}{\textbf{0.95 $\pm$ 0.02}}                                &                \textbf{0.95 $\pm$ 0.02}               \\ \hline

\multicolumn{6}{|c|}{\textbf{Abdominal CT }} \\ \hline
\multicolumn{1}{|c|}{\textbf{UNet}}  & 
\multicolumn{1}{c|}{0.77 $\pm$ 0.04}                               & 
\multicolumn{1}{c|}{0.77 $\pm$ 0.04}                               & 
\multicolumn{1}{c|}{0.76 $\pm$ 0.03}                                & 
\multicolumn{1}{c|}{0.74 $\pm$ 0.06}                                &             0.72   $\pm$ 0.06                  \\ \hline
\multicolumn{1}{|c|}{\textbf{VQ-UNet}} & 
\multicolumn{1}{c|}{\textbf{0.80 $\pm$ 0.01}}                               & 
\multicolumn{1}{c|}{\textbf{0.79 $\pm$ 0.01}}                               & 
\multicolumn{1}{c|}{\textbf{0.78 $\pm$ 0.01}}                                & 
\multicolumn{1}{c|}{\textbf{0.78 $\pm$ 0.02}}                                &           
\textbf{0.78     $\pm$ 0.02}                  \\ \hline

\multicolumn{6}{|c|}{\textbf{Prostate BMC dataset}}  \\ \hline

\multicolumn{1}{|c|}{\textbf{UNet}}  & 
\multicolumn{1}{c|}{0.80 $\pm$ 0.02}                               & 
\multicolumn{1}{c|}{0.80 $\pm$ 0.02}                               & 
\multicolumn{1}{c|}{0.76 $\pm$ 0.03}                                & 
\multicolumn{1}{c|}{0.75 $\pm$ 0.04}                                &     
0.70 $\pm$ 0.11                           \\ \hline
\multicolumn{1}{|c|}{\textbf{VQ-UNet}} & 
\multicolumn{1}{c|}{\textbf{0.82 $\pm$ 0.01}}                               & 
\multicolumn{1}{c|}{\textbf{0.82 $\pm$ 0.01}}                               &
\multicolumn{1}{c|}{\textbf{0.81 $\pm$ 0.02}}                                & 
\multicolumn{1}{c|}{\textbf{0.82 $\pm$ 0.01}}                                &        
\textbf{0.80 $\pm$ 0.02}                         \\ \hline

\end{tabular}
\end{table}

\begin{table}[]
\centering
\caption{Poisson noise perturbation results on all 3 datasets.}
\label{tab:PoissonResults}
\begin{tabular}{|cccccc|}
\hline
\multicolumn{1}{|c|}{}               & \multicolumn{1}{c|}{\textbf{Dice @0\% }} & \multicolumn{1}{c|}{\textbf{Dice @1\% }} & \multicolumn{1}{c|}{\textbf{Dice @10\% }} & \multicolumn{1}{c|}{\textbf{Dice @20\% }} & \textbf{Dice @30\% } \\ \hline
\multicolumn{6}{|c|}{\textbf{Chest X-ray NIH dataset}}  \\ \hline
\multicolumn{1}{|c|}{\textbf{UNet}}  & 
\multicolumn{1}{c|}{0.96 $\pm$ 0.02}                               & 
\multicolumn{1}{c|}{0.96 $\pm$ 0.02}                               & 
\multicolumn{1}{c|}{0.94 $\pm$ 0.04}                                & 
\multicolumn{1}{c|}{0.94 $\pm$ 0.04}                                &            
0.92     $\pm$ 0.05                 \\ \hline
\multicolumn{1}{|c|}{\textbf{VQ-UNet}} & 
\multicolumn{1}{c|}{\textbf{0.97 $\pm$ 0.01}}                               & 
\multicolumn{1}{c|}{\textbf{0.97 $\pm$ 0.01}}                               & 
\multicolumn{1}{c|}{\textbf{0.96 $\pm$ 0.01}}                                & 
\multicolumn{1}{c|}{\textbf{0.96 $\pm$ 0.02}}                                &              
\textbf{0.95 $\pm$ 0.02}                 \\ \hline

\multicolumn{6}{|c|}{\textbf{Abdominal CT}}  \\ \hline
\multicolumn{1}{|c|}{\textbf{UNet}}  & 
\multicolumn{1}{c|}{0.77 $\pm$ 0.01}                               & 
\multicolumn{1}{c|}{0.76 $\pm$ 0.02}                               & 
\multicolumn{1}{c|}{0.76 $\pm$ 0.02}                                & 
\multicolumn{1}{c|}{0.75 $\pm$ 0.02}                                &          
0.75 $\pm$ 0.02                       \\ \hline
\multicolumn{1}{|c|}{\textbf{VQ-UNet}} & 
\multicolumn{1}{c|}{\textbf{0.80 $\pm$ 0.01}}                               & 
\multicolumn{1}{c|}{\textbf{0.80 $\pm$ 0.01}}                               & 
\multicolumn{1}{c|}{\textbf{0.79 $\pm$ 0.01}}                                & 
\multicolumn{1}{c|}{\textbf{0.79 $\pm$ 0.02}}                                &           
\textbf{0.78 $\pm$ 0.02}                    \\ \hline

\multicolumn{6}{|c|}{\textbf{Prostate BMC dataset}}  \\ \hline
\multicolumn{1}{|c|}{\textbf{UNet}}  & 
\multicolumn{1}{c|}{0.80 $\pm$ 0.02}                               & 
\multicolumn{1}{c|}{0.80 $\pm$ 0.02}                               & 
\multicolumn{1}{c|}{0.81 $\pm$ 0.02}                                & 
\multicolumn{1}{c|}{0.78 $\pm$ 0.04}                                &      
0.77 $\pm$ 0.04                           \\ \hline
\multicolumn{1}{|c|}{\textbf{VQ-UNet}} & 
\multicolumn{1}{c|}{\textbf{0.82 $\pm$ 0.01}}                               & 
\multicolumn{1}{c|}{\textbf{0.81 $\pm$ 0.02}}                               & 
\multicolumn{1}{c|}{\textbf{0.82 $\pm$ 0.02}}                                & 
\multicolumn{1}{c|}{\textbf{0.82 $\pm$ 0.02}}                                &       
\textbf{0.80 $\pm$ 0.03}                         \\ \hline
\end{tabular}
\end{table}

\begin{figure}
    \centering
    \subfloat[NIH Chest X-ray sample]{\includegraphics[width=1\textwidth]{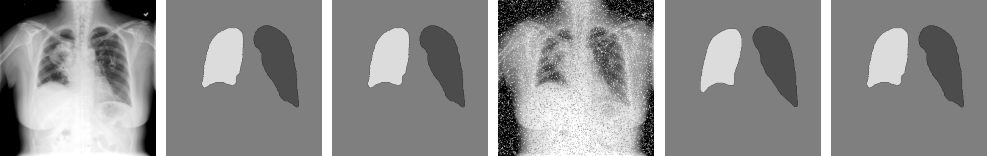}}\hfill
    \subfloat[BMC prostate sample]{\includegraphics[width=1\textwidth]{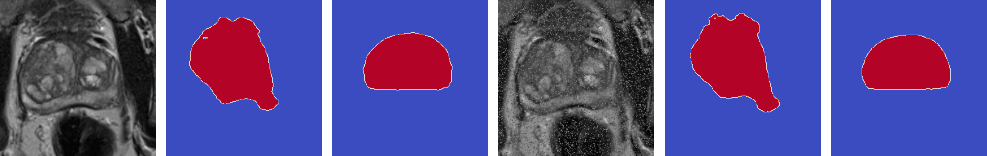}}\\
    \caption{Image and segmentation output for UNet and VQ-UNet under $0\%$ (1$^{st}$ 3 columns) and $30\%$ (2$^{nd}$ 3 columns) for s\&p noisee}
    \label{fig:noiserecon2}
\end{figure}
\begin{table}[]
\centering
\caption{Mean validation score for all 13 labels on the BTCV dataset}.
\label{tab:ABDCTresults}
\begin{tabular}{|ccccccccccccccc|}
\hline
\multicolumn{15}{|c|}{\textbf{Dice}}                                                                                                                                                                                                                                                    \\ \hline
\multicolumn{1}{|c|}{}               & \multicolumn{1}{c|}{\textbf{Spl}} & \multicolumn{1}{c|}{\textbf{Rki}} & \multicolumn{1}{c|}{\textbf{Lki}} & \multicolumn{1}{c|}{\textbf{Gal}} & \multicolumn{1}{c|}{\textbf{Eso}} &
\multicolumn{1}{c|}{\textbf{Liv}} & \multicolumn{1}{c|}{\textbf{Sto}} & \multicolumn{1}{c|}{\textbf{Aor}} & \multicolumn{1}{c|}{\textbf{IVC}} & \multicolumn{1}{c|}{\textbf{Eso}} &
\multicolumn{1}{c|}{\textbf{Vei}} & \multicolumn{1}{c|}{\textbf{Pan}} & \multicolumn{1}{c|}{\textbf{AG}} &
\multicolumn{1}{c|}{\textbf{Avg}} 
 \\ \hline
\multicolumn{1}{|c|}{\textbf{UNet}}  & \multicolumn{1}{c|}{0.94}                        
& \multicolumn{1}{c|}{0.86}                     
& \multicolumn{1}{c|}{0.86}                     
& \multicolumn{1}{c|}{0.63}
& \multicolumn{1}{c|}{0.73}
& \multicolumn{1}{c|}{0.94}                     
& \multicolumn{1}{c|}{0.84}                     
& \multicolumn{1}{c|}{0.81}
& \multicolumn{1}{c|}{\textbf{0.80}}                     
                     
& \multicolumn{1}{c|}{0.72}
& \multicolumn{1}{c|}{0.69}                     
& \multicolumn{1}{c|}{0.65}                     
& \multicolumn{1}{c|}{0.62} 
& \multicolumn{1}{c|}{0.77}                     
\\ \hline
\multicolumn{1}{|c|}{\textbf{VQ-UNet}}  & \multicolumn{1}{c|}{\textbf{0.95}}                        
& \multicolumn{1}{c|}{\textbf{0.88}}                     
& \multicolumn{1}{c|}{\textbf{0.89}}                     
& \multicolumn{1}{c|}{\textbf{0.63}}
& \multicolumn{1}{c|}{\textbf{0.76}}
& \multicolumn{1}{c|}{\textbf{0.95}}                     
& \multicolumn{1}{c|}{\textbf{0.84}}                     
& \multicolumn{1}{c|}{\textbf{0.84}}
& \multicolumn{1}{c|}{0.79}                     
                     
& \multicolumn{1}{c|}{\textbf{0.73}}
& \multicolumn{1}{c|}{\textbf{0.72}}                     
& \multicolumn{1}{c|}{\textbf{0.66}}                     
& \multicolumn{1}{c|}{\textbf{0.63}} 
& \multicolumn{1}{c|}{\textbf{0.79}}                

\\                                  \hline
\multicolumn{15}{|c|}{\textbf{95\% HD}}     \\ \hline                                     \multicolumn{1}{|c|}{\textbf{UNet}}  & \multicolumn{1}{c|}{\textbf{2.67}}                        
& \multicolumn{1}{c|}{2.98}                     
& \multicolumn{1}{c|}{2.95}                     
& \multicolumn{1}{c|}{\textbf{9.67}}
& \multicolumn{1}{c|}{4.91}                     
& \multicolumn{1}{c|}{2.80}                     
& \multicolumn{1}{c|}{3.31}
& \multicolumn{1}{c|}{5.12}                     
& \multicolumn{1}{c|}{8.19}                     
& \multicolumn{1}{c|}{\textbf{6.16}}
& \multicolumn{1}{c|}{8.10}                     
& \multicolumn{1}{c|}{7.82}                     
& \multicolumn{1}{c|}{7.01}
& \multicolumn{1}{c|}{5.51}

\\ \hline
\multicolumn{1}{|c|}{\textbf{VQ-UNet}}  
& \multicolumn{1}{c|}{2.69}                        
& \multicolumn{1}{c|}{\textbf{2.78}}                     
& \multicolumn{1}{c|}{\textbf{2.43}}                     
& \multicolumn{1}{c|}{9.81}
& \multicolumn{1}{c|}{\textbf{4.12}}                     
& \multicolumn{1}{c|}{\textbf{2.42}}                     
& \multicolumn{1}{c|}{\textbf{3.15}}
& \multicolumn{1}{c|}{\textbf{4.29}}                     
& \multicolumn{1}{c|}{\textbf{7.56}}                     
& \multicolumn{1}{c|}{6.20}
& \multicolumn{1}{c|}{\textbf{7.39}}                     
& \multicolumn{1}{c|}{\textbf{6.52}}                     
& \multicolumn{1}{c|}{\textbf{5.18}}
& \multicolumn{1}{c|}{\textbf{4.96}}

\\ \hline
\multicolumn{15}{|c|}{\textbf{ASD}}                                                                                                                                                                 
\\ \hline
\multicolumn{1}{|c|}{\textbf{UNet}}  & \multicolumn{1}{c|}{0.61}                        
& \multicolumn{1}{c|}{0.61}                     
& \multicolumn{1}{c|}{0.62}                     
& \multicolumn{1}{c|}{1.04}
& \multicolumn{1}{c|}{0.98}                     
& \multicolumn{1}{c|}{0.57}                     
& \multicolumn{1}{c|}{0.66}
& \multicolumn{1}{c|}{1.56}                     
& \multicolumn{1}{c|}{1.47}                     
& \multicolumn{1}{c|}{0.87}
& \multicolumn{1}{c|}{1.12}                     
& \multicolumn{1}{c|}{0.60}                     
& \multicolumn{1}{c|}{\textbf{0.61}}
& \multicolumn{1}{c|}{0.87}

\\ \hline
\multicolumn{1}{|c|}{\textbf{VQ-UNet}} &  \multicolumn{1}{c|}{\textbf{0.56}}                        
& \multicolumn{1}{c|}{\textbf{0.59}}                     
& \multicolumn{1}{c|}{\textbf{0.58}}                     
& \multicolumn{1}{c|}{\textbf{0.98}}
& \multicolumn{1}{c|}{\textbf{0.87}}                     
& \multicolumn{1}{c|}{\textbf{0.57}}                     
& \multicolumn{1}{c|}{\textbf{0.63}}
& \multicolumn{1}{c|}{\textbf{1.48}}                     
& \multicolumn{1}{c|}{\textbf{1.33}}                     
& \multicolumn{1}{c|}{\textbf{0.85}}
& \multicolumn{1}{c|}{\textbf{1.03}}                     
& \multicolumn{1}{c|}{\textbf{0.53}}                     
& \multicolumn{1}{c|}{0.61}
& \multicolumn{1}{c|}{\textbf{0.82}}                     
\\ \hline
\end{tabular}
\end{table}

\end{document}